\documentclass[10pt,twocolumn,letterpaper]{article}
\usepackage{multirow}
\usepackage{makecell} %
\usepackage{booktabs}       %
\usepackage{amsfonts}       %
\usepackage{nicefrac}       %
\usepackage{microtype}      %
\usepackage{graphicx}
\usepackage{amsmath} %
\usepackage{amssymb}  %
\usepackage{amsfonts}
\usepackage{mathtools}
\usepackage{commath}
\usepackage{color}
\usepackage{multirow}
\usepackage{makecell} %
\usepackage{colortbl}

\usepackage[normalem]{ulem}

\usepackage{caption}

\usepackage{pifont}   %
\usepackage{amsmath}
\usepackage{color,soul}

\usepackage[linesnumbered,ruled,vlined]{algorithm2e}
\usepackage{lipsum}
\usepackage[misc,geometry]{ifsym}

\usepackage[pagenumbers]{iccv} 

\definecolor{iccvblue}{rgb}{0.21,0.49,0.74}
\usepackage[pagebackref,breaklinks,colorlinks,allcolors=iccvblue]{hyperref}

\def\paperID{1634} 
\def\confName{ICCV}
\def\confYear{2025}

\newcommand{\MODELNAME}{MMAG}
\newcommand{\VAENAME}{MTA-VAE}
\newcommand{\ALLMODELNAME}{MAG}

\newcommand{\mysubsub}[1]{\noindent\textbf{#1}.}
\definecolor{c1}{HTML}{EF949E}
\definecolor{c2}{HTML}{FFD306}
\definecolor{c3}{HTML}{4874CB}

\newcommand{\tca}[1]{{\color{c1}#1}}
\newcommand{\tcb}[1]{{\color{c2}#1}}
\newcommand{\tccc}[1]{{\color{c3}#1}}

\title{
\tca{M}\tcb{A}\tccc{G}: \tca{M}ulti-Modal Aligned \tcb{A}utoregressive Co-Speech \\ \tccc{G}esture Generation without Vector Quantization}

\author{
Binjie Liu*$^{1,2}$ \quad Lina Liu*$^2$ \quad Sanyi Zhang$^{1\dagger}$ \\ \quad Songen Gu $^{3}$ \quad Yihao Zhi$^{4}$ \quad Tianyi Zhu$^{2}$ \quad Lei Yang$^{2}$ \quad Long Ye$^{1\dagger}$ \\
$^1$ Communication University of China \quad 
$^2$China Mobile Research Institute, Beijing, China \\
\quad $^3$School of Computer Science and Technology, University of \\
Chinese Academy of Sciences \quad $^4$The Chinese University of Hong Kong\\
}

\begin{document}

\twocolumn[{%
\renewcommand\twocolumn[1][]{#1}%
\maketitle
\centerline{
\includegraphics[width=0.82\linewidth]{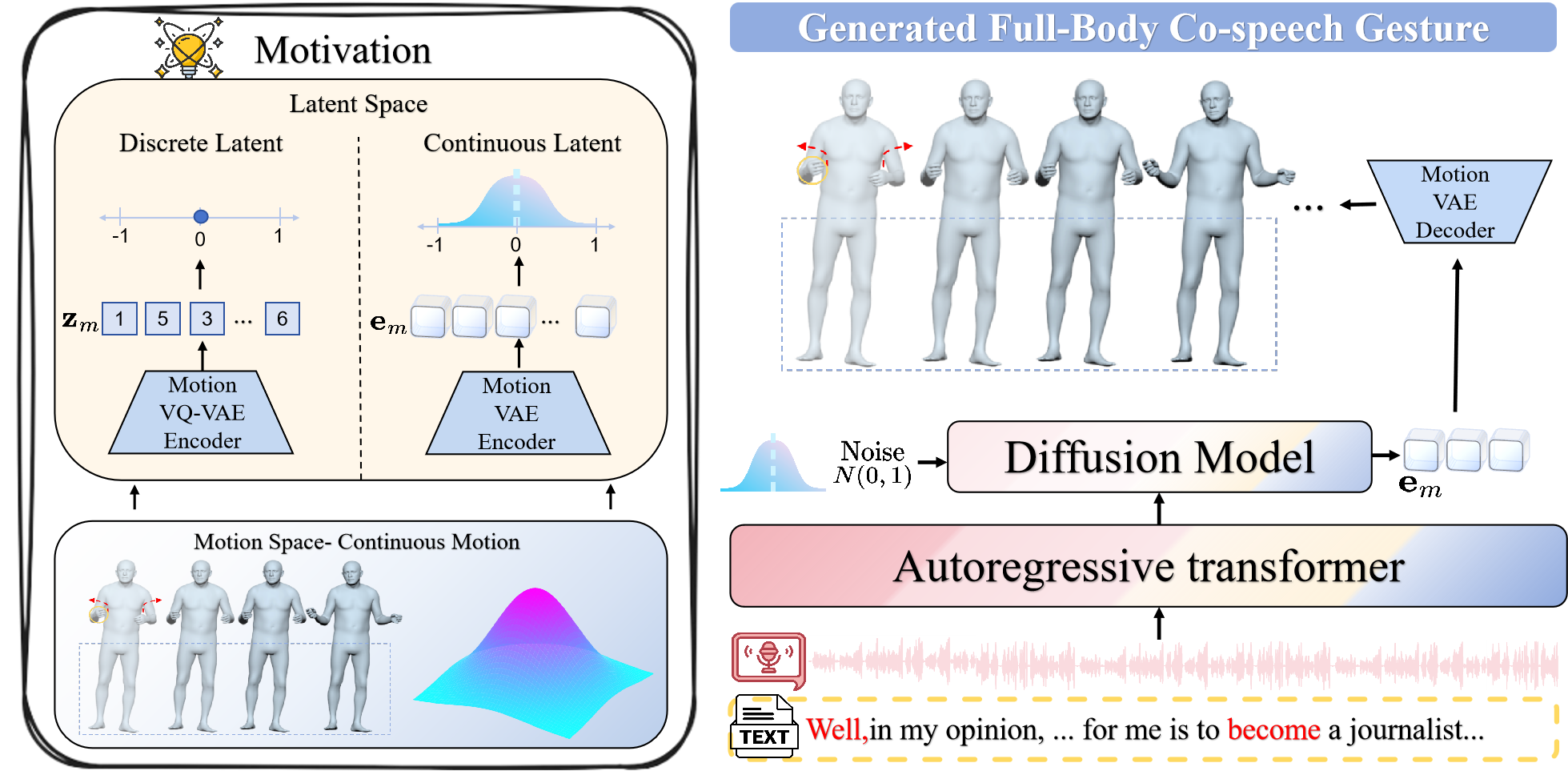}
}
\captionof{figure}{\textbf{On the left}, inspired by the natural continuity of human motion, we observe that VQ-VAE discretizes motion data, while VAE preserves a continuous latent space that better reflects real motion. \textbf{On the right}, motivated by this, we propose MAG, a framework that enables autoregressive modeling in continuous motion embeddings through diffusion, eliminating vector quantization. Given speech audio and text transcripts as conditionings, our model generates motion embeddings, which are decoded into realistic gestures via a Motion VAE decoder. \vspace{0.8em}}
\label{fig:teaser}
}]

\let\thefootnote\relax\footnotetext{$*$ Equal contribution.}
\let\thefootnote\relax\footnotetext{$\dagger$ Corresponding author.}

\begin{abstract}

This work focuses on full-body co-speech gesture generation. Existing methods typically employ an autoregressive model accompanied by vector-quantized tokens for gesture generation, which results in information loss and compromises the realism of the generated gestures. To address this, inspired by the natural continuity of real-world human motion, we propose \textbf{MAG}, a novel multi-modal aligned framework for high-quality and diverse co-speech gesture synthesis without relying on discrete tokenization. Specifically, (1) we introduce a motion-text-audio-aligned variational autoencoder (\textbf{MTA-VAE}), which leverages pre-trained WavCaps' text and audio embeddings to enhance both semantic and rhythmic alignment with motion, ultimately producing more realistic gestures. (2) Building on this, we propose a multi-modal masked autoregressive model (\textbf{MMAG}) that enables autoregressive modeling in continuous motion embeddings through diffusion without vector quantization. To further ensure multi-modal consistency, \textbf{MMAG} incorporates a hybrid granularity audio-text fusion block, which serves as conditioning for diffusion process. Extensive experiments on two benchmark datasets demonstrate that \textbf{MAG} achieves state-of-the-art performance both quantitatively and qualitatively, producing highly realistic and diverse co-speech gestures. The code will be released to facilitate future research.

\end{abstract}

\section{Introduction}


Full-body co-speech gesture generation aims to create realistic gestures that seamlessly complement verbal communication, which can serve as vital tools for conveying information in human communication~\cite{van1998persona,goldin2013gesture,cassell1999speech,hostetter2008visible,iverson1998people}. Automated generation of full-body co-speech gestures is a crucial technology for developing lifelike
avatars in diverse applications, such as virtual characters in the metaverse, human-computer interaction (HCI), and gaming to robot assistants~\cite{Yu2020SRG3SR,Rebol2021RealtimeGA}. 
However, there still exists two key challenges: 1) how to generate realistic and diverse gestures that seamlessly complement verbal communication; 2) how to generate semantically and rhythmically aligned gestures. 

Researchers have conducted substantial solutions in the field of full-body co-speech gesture generation. Traditional rule-based methods~\cite{10.1145/192161.192272, Cassell2004,Kipp2005GestureGB,huang2012robot, Kopp2002ModelbasedAO,Salem2009TowardsMR,10.1145/2485895.2485900} can integrate linguistic rules well, but the generated gestures lack flexibility and adaptability. In contrast, deep learning based approaches own favourable ability to handle complex multi-modal inputs. A wide range of strategies have been effectively explored for this task, including: normalizing flows~\cite{Ye2022AudioDrivenSG,Jonell2020LetsFI}, autoregressive models based on VQ-VAE~\cite{ao2022rhythmic,Ng2022LearningTL,Yi2022GeneratingH3,Liu2023EMAGETU,Xing2023CodeTalkerS3,Sun2024BeyondT,Chen2024EnablingSF,Ng2024FromAT,Chen2024TheLO}, GANs~\cite{Habibie2021LearningS3,Sadoughi2018NovelRO,Ginosar2019LearningIS,Ferstl2020AdversarialGG}, and diffusion models~\cite{alexanderson2022listen,Ao2023GestureDiffuCLIPGD,Yang2023DiffuseStyleGestureSA,Zhu2023TamingDM,Chen2024DiffSHEGAD}. This paper follows the learning-based solution upon diffusion model. 

Many existing learning-based approaches rely on autoregressive models with vector quantization (VQ) ~\cite{ao2022rhythmic,Ng2022LearningTL,Yi2022GeneratingH3,Liu2023EMAGETU,Xing2023CodeTalkerS3,Sun2024BeyondT,Chen2024TheLO} to tokenize motion data. However, as illustrated on the left side of ~\cref{fig:teaser}, real-world human motion usually exists in a continuous space. Discretizing it into categorical labels disrupts numerical magnitude relationships, leading to information loss and reducing the model’s ability to capture motion complexity. In addition, multi-modal information has not been effectively utilized to generate semantically and rhythmically aligned gestures. Methods such as MotionCLIP~\cite{tevet2022motionclip}, GestureDiffCLIP~\cite{Ao2023GestureDiffuCLIPGD}, and LivelySpeaker~\cite{Zhi2023LivelySpeakerTS} primarily focus on contrastive learning between text and motion, effectively capturing semantic content but neglecting the role of audio in rhythmic coordination. The joint alignment of motion with both text and audio embeddings remains an open challenge for generating more natural and synchronized gestures.

To tackle the above limitations, motivated by the continuity of real-world human motion, we propose \ALLMODELNAME, a multi-modal aligned autoregressive co-speech gesture generation framework, as illustrated on the right side of ~\cref{fig:teaser}. 
To maintain the high consistency between the multi-modal input and generated gestures, we simultaneously aggregate the motion, text, and audio into an unified learning module to enhance semantic and rhythmic coherence and generate more realistic gestures. Specifically, taking the motion, text and audio as input, 
we introduce a motion-text-audio-aligned variational autoencoder (MTA-VAE) supervised through contrastive learning, which leverages pre-trained WavCaps~\cite{Mei2023WavCapsAC} text and audio embeddings to enhance both semantic and rhythmic alignment with motion.
To meet the reliability distribution of real-world continuous motion space, we introduce to model with continuous motion embeddings by utilizing diffusion processes to avoid information loss in discretized representation. Specifically, we propose a multi-modal masked autoregressive model (MMAG) that facilitates autoregressive modeling without VQ through diffusion. Furthermore, a hybrid audio-text fusion block is incorporated by MMAG and served as conditioning for diffusion process. For final gesture generation, the continuous motion representations are predicted by diffusion network, and decoded by motion VAE decoder to generate more realistic gestures.

Our contributions are summarized as follows:

\begin{itemize}
    \item We propose a novel \textbf{\VAENAME}, the first to align motion, text and audio through contrastive learning to construct a unified multi-modal latent space, enabling accurate semantically aware and rhythmically aligned gesture generation.
    \item We introduce \textbf{\MODELNAME}, a multi-modal masked autoregressive gesture generation framework without VQ that enables autoregressive modeling in continuous motion embeddings through diffusion, avoiding information loss in discretized representation. 
    \item Experiments on two benchmark datasets confirm that \textbf{\ALLMODELNAME} achieves state-of-the-art performance. User studies further validate its effectiveness in generating realistic, synchronized, and diverse co-speech gestures.
\end{itemize}

\section{Related work}
\noindent\textbf{Realistic Co-speech Gesture Generation.}
As outlined above, co-speech gesture generation aims to produce realistic and diverse gestures that align with speech. Early rule-based methods~\cite{10.1145/192161.192272, Cassell2004, Kipp2005GestureGB, huang2012robot, Kopp2002ModelbasedAO, Salem2009TowardsMR, 10.1145/2485895.2485900} mapped linguistic rules to predefined gesture sequences but lacked flexibility. Deep learning has since become the dominant approach, with early neural models such as RNNs~\cite{Bhattacharya2021Speech2AffectiveGesturesSC} and CNNs~\cite{Pfister2014DeepCN} improving gesture synthesis by capturing temporal dependencies. More recently, masked autoregressive Transformers have been widely adopted for speech-aligned gesture generation. EMAGE~\cite{Liu2023EMAGETU} employs VQ-VAE to discretize body motion, enabling autoregressive Transformers to learn motion representations. However, these methods~\cite{ao2022rhythmic, Ng2022LearningTL, Yi2022GeneratingH3, Liu2023EMAGETU, Xing2023CodeTalkerS3, Sun2024BeyondT, Chen2024EnablingSF, Ng2024FromAT, Chen2024TheLO} rely on discrete tokenization techniques like VQ-VAE, which introduce quantization errors, ultimately degrading motion quality and diversity. To mitigate quantization errors, recent research has refined quantization techniques. ProbTalk~\cite{Liu2024TowardsVA} integrates PQ-VAE~\cite{Wu2018LearningPC} for holistic motion modeling, improving gesture representation through a two-stage refinement. Similarly, SynTalker~\cite{Ng2024FromAT} replaces VQ-VAE with RVQ-VAE to enhance precision and diversity. However, discretizing continuous motion data remains inherently lossy, posing a persistent challenge in co-speech gesture generation.

Many approaches utilize probabilistic models, including  normalizing flows~\cite{Ye2022AudioDrivenSG,Jonell2020LetsFI}, GANs~\cite{Habibie2021LearningS3,Sadoughi2018NovelRO,Ginosar2019LearningIS,Ferstl2020AdversarialGG} and diffusion models~\cite{alexanderson2022listen,Ao2023GestureDiffuCLIPGD,Yang2023DiffuseStyleGestureSA,Zhu2023TamingDM,Chen2024DiffSHEGAD}, for gesture generation. Diffusion models, in particular, refine noise into structured motion sequences through iterative denoising, demonstrating strong performance in human motion synthesis, especially in text-to-motion tasks. Their ability to generate diverse, high-quality gestures makes them well-suited for co-speech motion generation. For example, DiffuGesture~\cite{Zhao2023DiffuGestureGH} enhances audio-aligned gesture synthesis using classifier-free guidance and stabilization techniques, while DiffuseStyleGesture~\cite{Yang2023DiffuseStyleGestureSA} integrates cross-local and self-attention to ensure synchronization with speech. DiffSheg~\cite{Chen2024DiffSHEGAD} further advances this approach by introducing a diffusion-based Transformer with a unidirectional flow from expression to gesture, capturing joint distributions and employing an outpainting strategy for extended sequence generation.

Recent studies~\cite{Li2024AutoregressiveIG,Fan2024FluidSA,Yang2024MMARTL} in vision generation have shown that while both continuous and discrete tokenization adhere to the scaling law regarding validation loss, continuous tokenization outperforms discrete tokenization in terms of generation quality. This insight is particularly relevant in the context of co-speech gesture generation, which aims to produce realistic and diverse gestures that align with speech. To address the information loss caused by discretizing continuous motion data, we propose a novel framework that combines autoregressive modeling with diffusion models. Our approach enables autoregressive modeling directly in continuous motion embeddings through diffusion, leveraging the diffusion model's ability to generate high-quality motion data and the efficiency of Transformers in multimodal processing. This method improves the realism, diversity, and synchronization of generated co-speech gestures.

\noindent\textbf{Contrastive Learning in Co-speech Gesture Generation.}
Learning the alignment between audio, text, and motion is beneficial to generate realistic gestures. MotionCLIP~\cite{tevet2022motionclip} extends this by aligning motion data with the pretrained CLIP text space. GestureDiffCLIP~\cite{Ao2023GestureDiffuCLIPGD} and LivelySpeaker~\cite{Zhi2023LivelySpeakerTS} introduce a gesture-transcript alignment mechanism using contrastive learning. Similarly, LivelySpeaker~\cite{Zhi2023LivelySpeakerTS}, an MLP-based diffusion model, leverages the pretrained CLIP model to generate semantically aligned gestures.

However, performing contrastive learning between text-only features and gesture features is challenging due to the absence of timing information. Additionally, generating gestures solely from audio struggles to capture the semantic relationships. To address these challenges, we propose using contrastive learning within a VAE framework to align the audio, text, and gesture modalities. This approach enhances gesture generation by improving both semantic alignment and rhythm synchronization.

\section{Method}
\begin{figure*}[htbp]
    \centering
    \includegraphics[width=0.9\linewidth ]{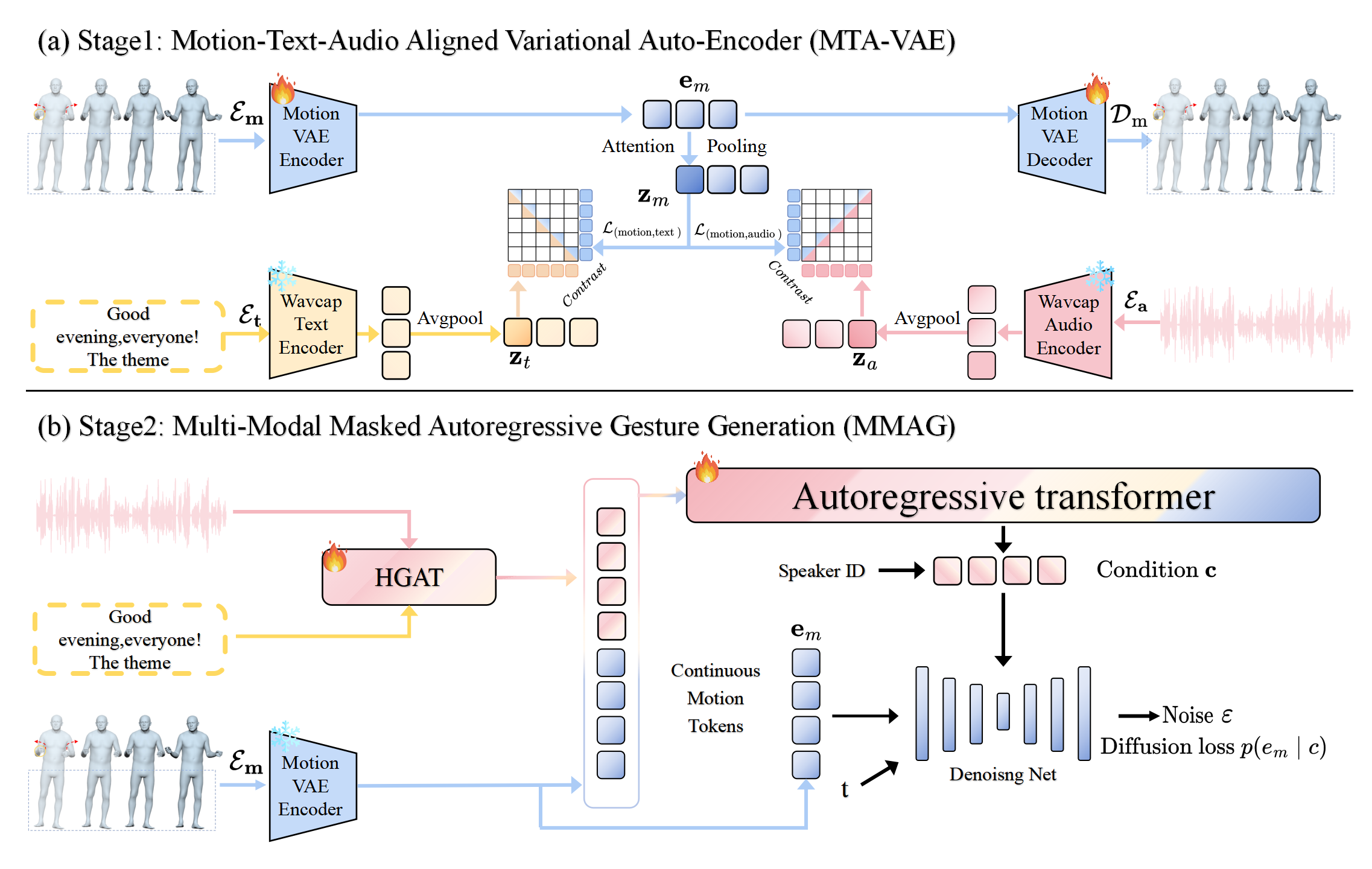}
    \caption{\textbf{Architecture of MAG.} MAG generates realistic co-speech gestures in two stages: (a) \VAENAME: Motion VAE encodes motion embeddings $\mathbf{e}_m$, which are aligned with WavCaps' text embeddings $\mathbf{z}_t$ and audio embeddings $\mathbf{z}_a$ through contrastive learning. (b) \MODELNAME: MMAG utilizes an autoregressive model to predict each motion embedding, derived from MTA-VAE encoder.The diffusion process, which predicts noise and ultimately generates the motion embedding, is guided by a text-audio fusion mechanism within a hybrid granularity audio-text fusion block. This ensures coherence across modalities, with the final motion output being produced by the MTA-VAE decoder.}
    \label{fig:pipeline.png}
\end{figure*}
\vspace{-0.4em}

\setlength{\belowcaptionskip}{-0.05cm}
\begin{figure}[htbp]
\centerline{
\includegraphics[width=0.9\columnwidth]{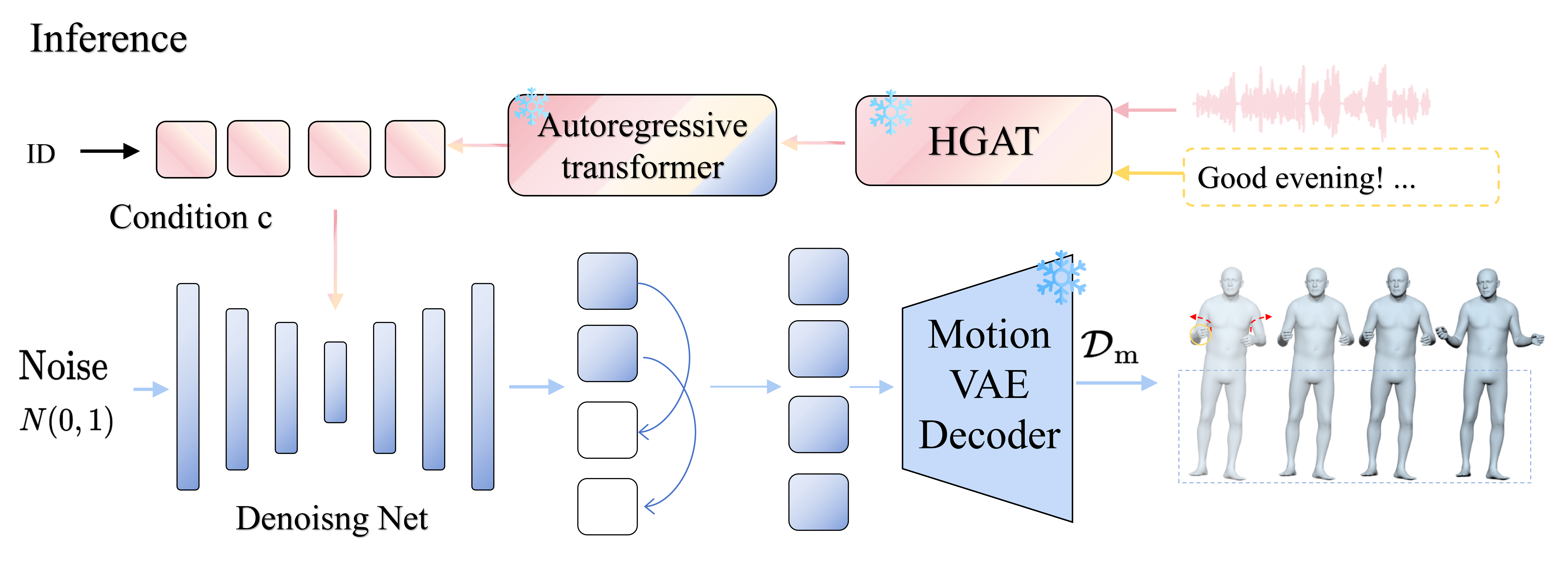}
}
\vspace{-0.02in}
\caption{
\textbf{Inference.}
By feeding the noise and conditioning $\mathbf{c}$ from the MMAG input into the denoising network, we can accurately generate the motion embeddings, which are then reconstructed into the real motion by the motion VAE decoder. 
}
\vspace{-0.8em}
\label{fig:inference}
\end{figure}

\subsection{Overview}

The keys of our proposed framework are continuous embedding representations and the joint alignment of motion with both text and audio embeddings to generate high-quality and diverse gestures. To achieve this, as shown in Fig.~\ref{fig:pipeline.png}, our framework consists of two stages: the Motion-Text-Audio-Aligned Variational Auto-Encoder (\VAENAME) and the Multi-modal Masked Autoregressive Gesture Generation Model (\MODELNAME). See \cref{sec3:\VAENAME} and \cref{sec3:MMAG} for details.

\mysubsub{Training}
For the first stage, to generate semantically and rhythmically aligned continuous motion embeddings, motion, text and audio are input to MTA-VAE. Motion is encoded and decoded by motion VAE encoder and decoder to get continuous motion embeddings. Text and audio are encoded to get text and audio features, and motion embeddings are aligned to these features through contrastive learning. Then the learned motion VAE and continuous motion embeddings are fed to the next stage. The goal of the second stage is to transform the learned motion embeddings to noise using diffusion process. MMAG use the learned motion VAE encoder to predict motion embeddings to input to denoising network. Furthermore, the hybrid granularity audio-text fusion block is incorporated by MMAG, and the fused features are fed to an autoregressive transformer to get conditioning used for diffusion process.

\mysubsub{Inference}
As shown in Fig.~\ref{fig:inference}, taking the noise and conditioning $c$ predicted from text, audio and speaker ID as input, denoising network can predict continuous motion embeddings. Then, the continuous motion embeddings are input to motion VAE decoder to generate final gestures.

\subsection{Motion-Text-Audio-Aligned Variational Auto-Encoder}
\label{sec3:\VAENAME}

To generate semantically and rhythmically aligned gestures, different from previous works~\cite{tevet2022motionclip, Ao2023GestureDiffuCLIPGD, Zhi2023LivelySpeakerTS} which focus on aligning text and motion to improve semantic awareness, we propose the Motion-Text-Audio-Aligned Variational Autoencoder (MTA-VAE) to utilize the stronger correlation between audio and motion, as illustrated in Fig.\ref{fig:pipeline.png} (a). Unlike previous approaches that rely on discrete vector-quantized motion tokenization and contrastive learning only between text and motion, MTA-VAE integrates a motion VAE (encoder and decoder) with continuous motion representation, and contrastive learning in motion, text and audio. By leveraging continuous representations, our method reconstructs multi-modal aligned motion sequences, effectively capturing intricate motion details while enhancing the coherence between audio, text, and motion.

Following EMAGE~\cite{Liu2023EMAGETU}, we divide body gestures into three components— upper body, hands, and lower body—represented as $\mathbf{M} = \{\mathbf{m}_\text{u}, \mathbf{m}_\text{h}, \mathbf{m}_\text{l}\}$. Each component is fed into MTA-VAE to learn continuous motion representation, separately. For simplicity, we use a unified $\mathbf{m}$ to represent one of the components, and finally the learned three components are combined and input to the next stage.

\mysubsub{Motion VAE with Continuous Motion Representation}
\label{sec:Continuous Motion Representation.}
The motion VAE consists of motion VAE encoder $\mathcal{E}_{\mathbf{m}}$ and motion VAE decoder $\mathcal{D}_{\mathbf{m}}$. Taking motion sequence $\mathbf{m}$ with $T$ frames as input, $\mathcal{E}_{\mathbf{m}}$ processes motion sequence $\mathbf{m}$ using a four-layer Temporal Convolutional Network (TCN)~\cite{Bai2018AnEE}, extracting continuous motion features $\mathbf{e}_{\mathbf{m}} \in \mathbb{R}^{T\times64}$. $\mathcal{D}_{\mathbf{m}}$ then reconstructs the motion sequence from these continuous features. The above process can be formulated as: 
\begin{equation}
\begin{aligned}
    \mathbf{e}_{\mathbf{m}} = \mathcal{E}_{\mathbf{m}}(\mathbf{m}) \ ,\  \hat{\mathbf{m}} = \mathcal{D}_{\mathbf{m}}(\mathbf{e}_{\mathbf{m}})
\end{aligned}
\end{equation}

where $\mathbf{\hat{m}}$ represents the reconstructed motion. The reconstructed $\hat{\mathbf{m}}$ is supervised by $\mathbf{m}$ using motion loss $\mathcal{L}_{\text{motion}}$:
\begin{align}
\begin{split}
    \mathcal{L}_{\text{motion}} =& 
    \mathcal{L}_{\text{rec}} (\mathbf{m}, \mathbf{\hat{m}}) + 
    \mathcal{L}_{\text{vel}} (\mathbf{m}', \mathbf{\hat{m}'}) + \\
    & 
    \mathcal{L}_{\text{acc}} (\mathbf{m}'', \mathbf{\hat{m}''})
\end{split}
\end{align}

where $\mathbf{m}'$ and $\mathbf{\hat{m}'}$ denote the velocity of $\mathbf{m}$ and $\mathbf{\hat{m}}$, respectively, and $\mathbf{m}''$ and $\mathbf{\hat{m}''}$ represent their acceleration. $\mathcal{L}_{\text{rec}}$ is computed using Geodesic~\cite{Tykkala2011DirectIC}, while $\mathcal{L}_{\text{vel}}$ and $\mathcal{L}_{\text{acc}}$ are computed using L1 losses.

\mysubsub{Contrastive Learning in Motion,Text and Audio}
\label{sec:Contrast Learning in Gesture,Text and Audio.}
Given an input triplet of motion $\mathbf{m}$, text $\mathbf{t}$, and audio $\mathbf{a}$, the text and audio are utilized to supervise the learning of motion embeddings through contrastive learning. For better generalization, we leverage WavCaps~\cite{Mei2023WavCapsAC}, which is trained on a large-scale audio-text dataset containing approximately 400k audio clips with paired text, to extract text and audio features. Specifically, the frozen audio features $\mathbf{z}_a$ are extracted from WavCaps' HTSAT encoder $\mathcal{E}_\mathbf{a}$~\cite{Chen2022HTSATAH}, text features $\mathbf{z}_t$ are from WavCaps' BERT-based text encoder $\mathcal{E}_\mathbf{t}$~\cite{devlin2018bert}, and motion embeddings $\mathbf{e}_m$ are produced by the motion VAE encoder $\mathcal{E}_\mathbf{m}$. 
The motion embeddings \(\mathbf{e}_m \in \mathbb{R}^{T \times 64}\) are reshaped by a linear projection layer to \(\mathbf{z}_m \in \mathbb{R}^{T \times 1024}\), aligning them with the unified text-audio feature space. They are then further aggregated into the most representative features \(\mathbf{z}_m \in \mathbb{R}^{1 \times 1024}\) through an attention pooling layer~\cite{Ye2023CLAPSpeechLP}. The objective of contrastive learning is to maximize the similarity of the positive pairs. The contrastive loss $\mathcal{L}_{\text{m,t,a}}$ is defined as:

\begin{equation}
\begin{aligned}
    \mathcal{L}_{(m2a)} &=  - \frac{1}{2} \sum\limits_{(j,k)} 
    \log \frac{\exp (\mathbf{z}_m^j \mathbf{z}_a^k) }{ \sum\limits_l \exp (\mathbf{z}_m^j \mathbf{z}_a^l) },\\ 
    \mathcal{L}_{(a2m)} &=  - \frac{1}{2} \sum\limits_{(j,k)} 
    \log \frac{\exp (\mathbf{z}_m^j \mathbf{z}_a^k) }{ \sum\limits_l \exp (\mathbf{z}_m^l \mathbf{z}_a^k)},\\
    \mathcal{L}_{(m2t)} &= - \frac{1}{2} \sum\limits_{(j,k)}
    \log \frac{\exp (\mathbf{z}_m^j \mathbf{z}_t^k) }{ \sum\limits_l \exp (\mathbf{z}_m^j \mathbf{z}_t^l) },\\
    \mathcal{L}_{(t2m)}  &= - \frac{1}{2} \sum\limits_{(j,k)}
    \log \frac{\exp (\mathbf{z}_s^j \mathbf{z}_t^k) }{ \sum\limits_l \exp (\mathbf{z}_m^l \mathbf{z}_t^k) },\\
    \mathcal{L}_{\text{m,t,a}} =& \mathcal{L}_{(m2a)}+\mathcal{L}_{(a2m)}+\mathcal{L}_{(m2t)}+\mathcal{L}_{(t2m)},
\end{aligned}
\end{equation}
where $(j, k)$ indicates the positive pair in training batches. Since the audio-text alignment is handled by the pre-trained WavCaps encoders, there is no need to caculate $\mathcal{L}_{(t2a)}$ and $\mathcal{L}_{(a2t)}$.  

The total loss for training \VAENAME{ } is formulated as:
\begin{equation}
\begin{aligned}
    \mathcal{L}_{\text{total}} =& \mathcal{L}_{\text{motion}}+ \lambda_{c} \mathcal{L}_{\text{m,t,a}} +  \lambda_{KL} \mathcal{L}_{\text{kl}},
\end{aligned}
\end{equation}
where $\mathcal{L}_{\text{motion}}$ represents the motion loss, $\mathcal{L}_{\text{m,t,a}}$ corresponds to the contrastive loss, and $\mathcal{L}_{\text{kl}}$ denote Kullback–Leibler(KL)~\cite{kl} divergence which regularizes the latent embedding $\mathbf{e}_m$ to follow Gaussian prior.

\subsection{Multi-modal Masked Autoregressive Gesture Generation}
\label{sec3:MMAG}
Multi-modal masked autoregressive gesture generation (MMAG) use the learned motion VAE encoder to predict motion embeddings to input to denoising net, as illustrated in Fig.~\ref{fig:pipeline.png} (b). The hybrid granularity audio-text fusion block (HGAT) is incorporated by MMAG, and the fused features are fed to an autoregressive transformer to get conditioning, which is used as the conditioning for diffusion process. Furthermore, identity enoding is incorporated to MMAG to enable personalized aligned gestures.

\mysubsub{Identity Encoding}
\label{sec3:Identity Encoding.}  
Following EMAGE \cite{Liu2023EMAGETU}, we use one-hot encoding for speaker identity. Incorporating speaker ID enhances the model's ability to capture individual speaking styles, enabling personalized aligned gestures.

\setlength{\belowcaptionskip}{-0.5cm}
\begin{figure}[htbp]
\centerline{
\includegraphics[width=0.9\columnwidth]{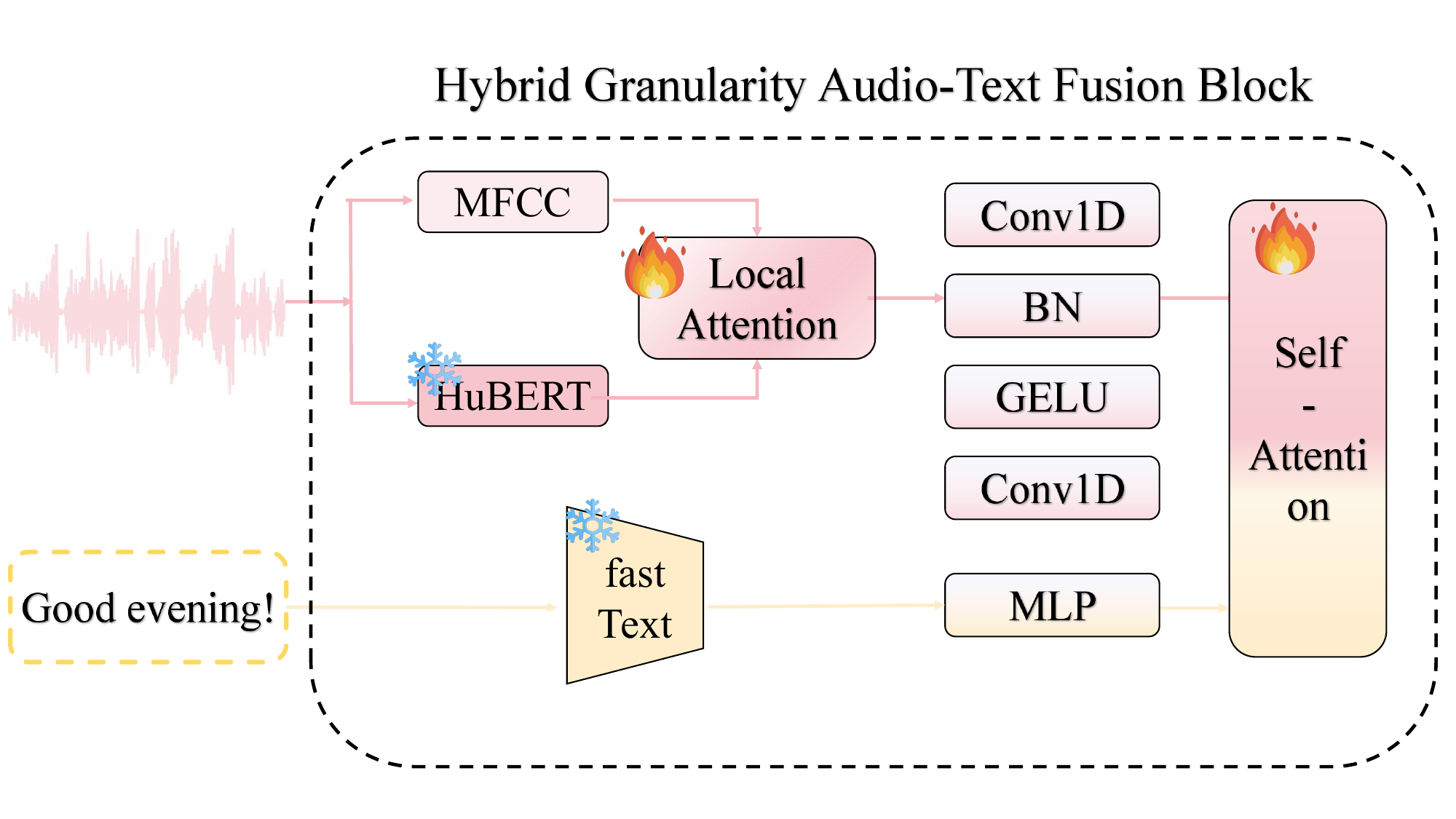}
}
\vspace{-0.03in}
\caption{
\textbf{Architecture of HGAT.}
The Hybrid Granularity Audio-Text Fusion Block processes audio and text inputs, extracting low-level audio features (MFCC) for rhythm synchronization and high-level features (HuBERT) for semantic understanding, while text features are extracted using fastText. These features are then fused through attention mechanisms .
}
\label{fig:HGAT}
\end{figure}
\vspace{-0.01em}

\mysubsub{Hybrid Granularity Audio-Text Fusion Block}
\label{sec3:HGAT.}  
Gestures are inherently influenced by both text semantics and audio rhythm. To achieve audio-text fusion, we introduce the Hybrid Granularity Audio-Text Fusion Block (HGAT) (see Figure \ref{fig:HGAT}).
For audio, we extract and integrate features at two levels: low-level and high-level. The low-level features are derived from Mel-Frequency Cepstral Coefficient (MFCC), which capture fundamental acoustic properties, emphasizing the low-frequency information that is typically more abundant in human speech, ensuring synchronization with speech rhythm. In contrast, high-level features are from HuBERT~\cite{Hsu2021HuBERTSS}, which encode richer semantic information, enabling better recognition of emotional cues within speech. Then, these two audio features are fused using local attention~\cite{Roy2020EfficientCS}.
For text, we extract features using fastText \cite{Bojanowski2016EnrichingWV}. To enhance audio-text integration, we downscale audio features and upscale text features before merging them through self-attention. This enables a seamless multi-modal representation, blending rhythm and content for more synchronized and expressive gestures.


\mysubsub{Masked Autoregressive Gesture Generation Model}
\label{sec3:Multi-modal Masked Autoregressive Gesture Generation Model.}  
Inspired by MAR~\cite{Li2024AutoregressiveIG}, we introduce MMAG, a multimodal masked autoregressive gesture generation model that integrates the advantages of masked autoregressive transformers and diffusion models for latent motion embedding generation.
Unlike previous methods that rely on VQ tokenizers, we represent continuous motion embeddings as tokens, preserving motion continuity and eliminating information loss caused by quantization.
In MMAG, motion, text and audio are utilized for multi-modal learning, where motion is used to generate continuous motion embeddings, motion, text and audio are fused and fed into an autoregressive transformer to generate conditioning. And we employ a diffusion-based MLP denoising network~\cite{He2015DeepRL} to receive these embeddings and conditioning. 
Specifically, the three \VAENAME{} encoders processes known motion sequence data \(\mathbf{M}\) (hands,upper body, and lower body) to generate continuous motion embeddings \(\mathbf{e}_M = \{\mathbf{e}_M^1, ..., \mathbf{e}_M^i, ..., \mathbf{e}_M^T\}\in \mathbb{R}^{T \times 64 \times 3}\), where \(1 \leq i \leq T\), $T$ is the number of frames.
To improve prediction accuracy, the text-audio fusion embeddings from HGAT are applied to guide the generation of the next motion embedding based on known embeddings. Upon receiving the text-audio fusion input, the autoregressive Transformer uses a network $c = f(\cdot)$ to generate $\mathbf{c}$ which serves as the conditioning input for the denoising network.
Furthermore, the probability of the next embedding is modeled as $p(\mathbf{e}_M|\mathbf{c})$. We use Eqn.(\ref{eq:denoise2}) as the loss function to model the per-embedding probability.
\begin{equation}
\mathcal{L}(c, \mathbf{e}_M)=\mathbb{E}_{\varepsilon, t}\left[\left\|\varepsilon-\mathcal{E}_\theta\left(\mathbf{e}_{M,t} \mid t, c\right)\right\|^2\right],
\label{eq:denoise2}
\end{equation}
where \(\varepsilon \sim \mathcal{N}(\mathbf{0}, \mathbf{I})\) is Gaussian noise, $t$ is a time step of the noise schedule. $\mathcal{E}_\theta$ represents the noise estimator, which is our MLP denoising network parameterized by $\theta$. The noisy motion latent token \(\mathbf{e}_{M,t}\) is:  
\begin{equation}
\mathbf{e}_{M,t} = \sqrt{\bar{\alpha}_t} \mathbf{e}_M + \sqrt{1-\bar{\alpha}_t} \varepsilon.
\end{equation}
where $\bar{\alpha}_t$ defines a noise schedule \cite{Ho2020DenoisingDP}. And the goal of the MLP denoising network is to generate continuous-valued motion embeddings in autoregressive mode. Therefore, the final trained MLP denoising network takes noise as input and, conditioned on $\mathbf{c}$ and $t$, predicts continuous-valued motion embeddings. These embeddingss are subsequently reconstructed into real motion data using the motion VAE decoder.

\section{Experiments}
\begin{figure*}[htbp]
    \centering
    \includegraphics[width=0.88\linewidth ]{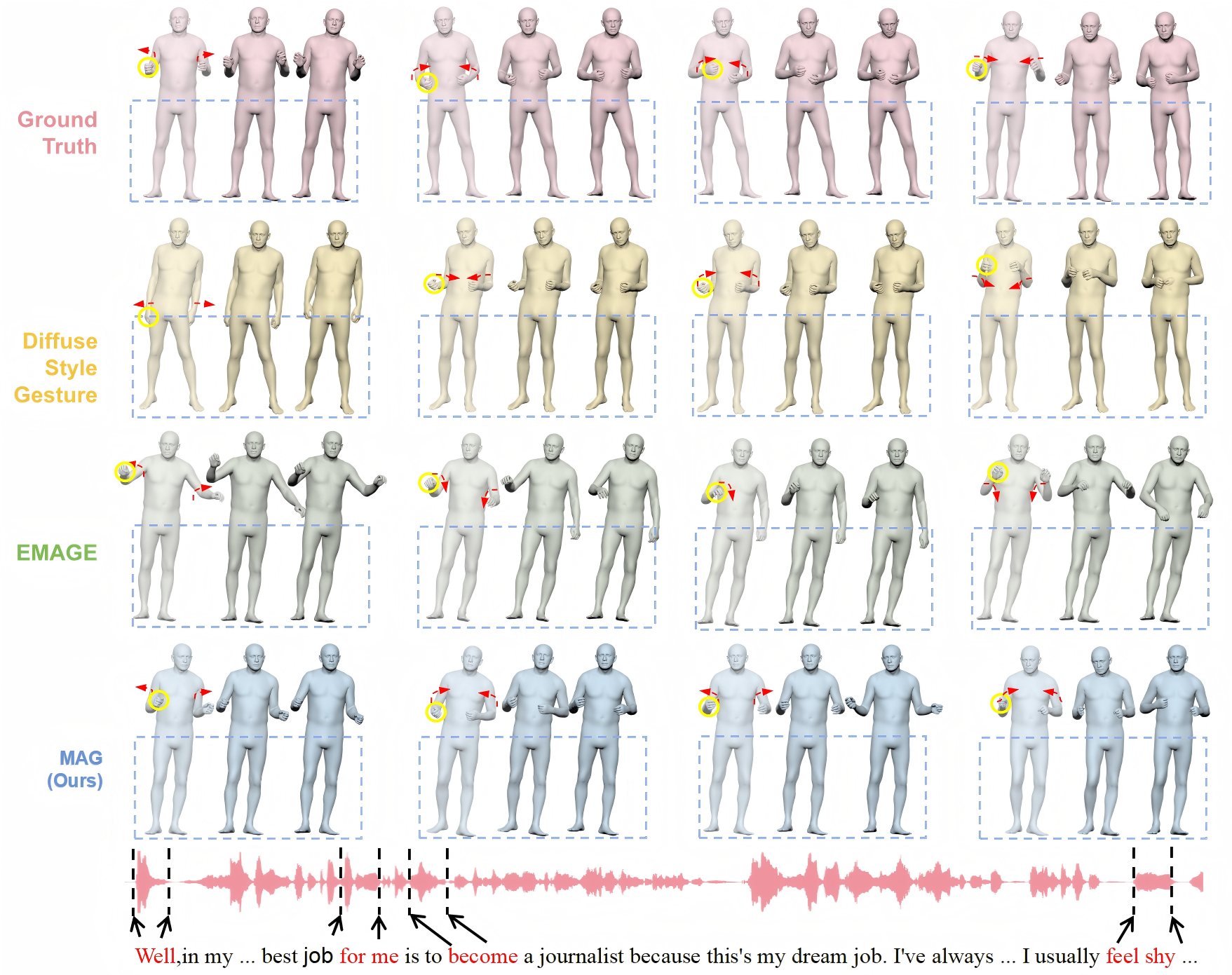}
    \caption{
    \textbf{Qualitative Comparison on BEATv2 Dataset.}
     Compared to other methods, our MAG approach generates gestures that more closely resemble GroundTruth and achieve better synchronization with both audio and text input. The gestures respond to high-frequency segments in the audio and show semantically relevant movements for meaningful words such as "well," "become," "for me," and "feel shy."
    }
    \label{fig:contrast.png}
\end{figure*}

\begin{figure*}[htbp]
    \centering
    \includegraphics[width=0.9\linewidth ]{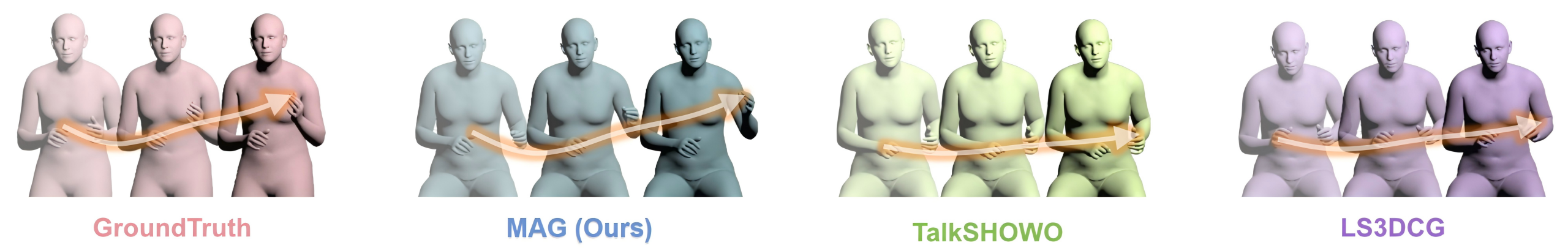}
    \caption{
    \textbf{Qualitative Comparison on Show Dataset.}
     Compared to other methods, our MAG approach generates gestures that more closely resemble GroundTruth or other similar dynamic actions when dealing with speech containing strong semantic information, such as words like "this" and "mistake".
    }
    \label{fig:contrast2.png}
    \vspace{-1em}
\end{figure*}

\subsection{Experimental Setup} 
\mysubsub{Datasets} 
BEATv2, introduced in EMAGE~\cite{Liu2023EMAGETU}, is a large-scale multimodal human gesture dataset containing text transcripts, semantic and emotional labels, and 60 hours of motion data. ~\cite{Liu2023EMAGETU} splits it into BEAT2-Standard (27 hours) and BEAT2-Additional (33 hours). Following the same settings as in ~\cite{Liu2023EMAGETU}, we report results on the BEAT2-Standard Speaker-2 subset with an 85\%/7.5\%/7.5\% for training, validation, and testing, respectively.

SHOW~\cite{talkshow:yi2022generating} is an audio-visual dataset featuring 3D full-body mesh annotations synchronized with audio from in-the-wild talk show recordings. It includes 26.9 hours of talk show footage from four speakers, with SMPLX~\cite{Pavlakos2019ExpressiveBC} parameters at 30 fps and corresponding audio sampled at 22 kHz. Following ~\cite{talkshow:yi2022generating}, we select video sequences longer than three seconds and split them into 80\%/10\%/10\% for training, validation, and testing, respectively.

\mysubsub{Evaluation Metrics} 
Following previous methods~\cite{Liu2023EMAGETU}, we adopt three key metrics to evaluate the quality of body gesture generation. Frechet Gesture Distance (FGD)~\cite{trimodal} measures the distribution disparity between generated and ground-truth gestures (GT). Beat Consistency Score (BC)~\cite{ha2g:liu2022learning} assesses the synchronization between speech and motion. Diversity~\cite{ha2g:liu2022learning} is calculated as the average L1 distance across multiple body gesture clips, reflecting the variability of generated motions.

\mysubsub{Implementation Details}
We train our model on a single NVIDIA A800  GPUs. For \textbf{BEATv2} dataset, we train our model for 800 epochs with a batch size of 120. For \textbf{SHOW} dataset, we train 30 epochs with a batch size of 120.

\subsection{Quantitative Results}
\mysubsub{Quantitative Comparisons}
On the BEATv2 dataset, we follow the evaluation protocol of ~\cite{Liu2023EMAGETU} to ensure a fair comparison with baseline methods~\cite{ginosar2019learning,yoon2020speech,ha2g:liu2022learning,liu2022disco,Yang2023DiffuseStyleGestureSA,Habibie2021LearningS3,talkshow:yi2022generating,Liu2023EMAGETU,Chen2024TheLO}. As shown in Table~\ref{tab:tab1}, MAG outperforms existing approaches in FGD, indicating that its generated gestures are closest to the ground truth. MAG also maintains a high BC score, ensuring strong speech-gesture synchronization. While TheLO~\cite{Chen2024TheLO} and DiffStyleGesture~\cite{Yang2023DiffuseStyleGestureSA} achieve competitive diversity scores, MAG lowers FGD while maintaining a relatively high diversity score. MAG strikes a better balance between motion realism, synchronization, and diversity.

On the BEATv2 dataset, we follow the evaluation protocol of ~\cite{Liu2023EMAGETU} to ensure a fair comparison with baseline methods~\cite{ginosar2019learning,yoon2020speech,ha2g:liu2022learning,liu2022disco,Yang2023DiffuseStyleGestureSA,Habibie2021LearningS3,talkshow:yi2022generating,Liu2023EMAGETU,Chen2024TheLO}. As shown in Tab.~\ref{tab:tab1}, \textbf{MAG} surpasses existing approaches in \textbf{FGD}, demonstrating that its generated gestures are the closest to the ground truth. It also maintains highest \textbf{BC} score, ensuring strong speech-gesture synchronization. While TheLO~\cite{Chen2024TheLO} and DiffStyleGesture~\cite{Yang2023DiffuseStyleGestureSA} achieve competitive diversity scores, \ALLMODELNAME{ }effectively reduces diversity while preserving a high level of FGD. Overall, \ALLMODELNAME{ }achieves a well-balanced trade-off between motion realism, synchronization, and diversity, setting a new benchmark for co-speech gesture generation.

On the SHOW dataset, we follow to the TalkSHOW~\cite{talkshow:yi2022generating} evaluation setting and focus on body gesture generation, comparing MAG with TalkSHOW and LS3DCG~\cite{Habibie2021LearningS3}. All baseline methods generate gestures by modeling only upper-body and hand movements. As shown in Tab.~\ref{tab:tab1}, \textbf{\ALLMODELNAME}{ }achieves superior results in \textbf{FGD}, indicating that its generated gestures closely resemble real human motion. Additionally, it achieves the high \textbf{Diversity} score, demonstrating its ability to generate more expressive gestures while maintaining motion fidelity.

\begin{table}[htbp]
    \centering
    \resizebox{0.99\linewidth}{!}{
        \begin{tabular}{l|c|ccccc}
        \toprule
                    Method& Dataset   & FGD $\downarrow$ & BC $\uparrow$ & Diversity~$\uparrow$  \\ 
        \midrule
        S2G\cite{ginosar2019learning}   &\multirow{12}{*}{BEATv2}          & 28.15                 & 4.68      & 5.97                        \\
        Trimodal\cite{yoon2020speech}     &    & 12.41                 & 5.93                & 7.72                       \\
        HA2G\cite{ha2g:liu2022learning}   &          & 12.32                 & 6.78               & 8.62                       \\
        DisCo\cite{liu2022disco}  &          & 9.417                 & 6.44                & 9.91                       \\
        CaMN\cite{liu2022beat}   &          & 6.644                  & 6.77               & 10.86                        \\
        DiffStyleGesture\cite{Yang2023DiffuseStyleGestureSA} && 8.811                  & 7.24                & 11.49                      \\
        Habibie \textit{et al}.\cite{Habibie2021LearningS3}  & & 9.040                & 7.71                & 8.21                    \\
        DiffStyleGesture \cite{talkshow:yi2022generating} &        & 6.209                  & 6.94                & 13.47                  \\
        EMAGE\cite{Liu2023EMAGETU}  &    & 5.512                  & 7.72               & 13.06                    \\
        TheLO\cite{Chen2024TheLO} &       & \underline{5.30}                  & \underline{7.78}                & \underline{15.16}                  \\
        \rowcolor{gray!20}
        \ALLMODELNAME{ }(VAE) &     & \textbf{4.835}                  & \textbf{7.84}                & 12.85             \\
        \rowcolor{gray!20}
        \ALLMODELNAME{ }(\VAENAME) &     & \textbf{4.565}                  & \textbf{7.84}                & \textbf{13.27}            \\
        \midrule
        LS3DCG\cite{Habibie2021LearningS3} &\multirow{3}{*}{Show}  & 2392.3    & \textbf{9.477}      &  \underline{4.539}              \\
        TalkSHOW\cite{talkshow:yi2022generating} &  &\underline{1155.6}      & \underline{8.70}      & 4.365 \\
        \rowcolor{gray!20}
        \ALLMODELNAME{ }(VAE)  &    & \textbf{592.7}                  & 8.28                & \textbf{5.190}           \\
        \midrule
    \end{tabular}
    }
    \vspace{-0.1in}
    
    \caption{\textbf{Quantitative evaluation on BEATv2 and Show.} The "$\uparrow$" indicates the higher, the better and "$\downarrow$" means the lower, the better. \textbf{Bold} and \underline{underline} represent optimal and suboptimal results. We report FGD $\times 10^{-1}$, BC $\times 10^{-1}$ and Diversity as EMAGE.}
    \label{tab:tab1}
\end{table}

\begin{table}[htbp]
    \vspace{0.08in}

    \centering
    \resizebox{1\linewidth}{!}{
        \begin{tabular}{l|c|cccc}
        \toprule
                    Method& Dataset  & Real &  Div & G-S Sync & Sem  \\ 
        \midrule
        DiffStyleGesture \cite{talkshow:yi2022generating}  &\multirow{3}{*}{BEATv2}     & 2.364      & 2.718      & 2.541      & 2.687           \\
        EMAGE\cite{Liu2023EMAGETU}  &    & 1.989     & 1.750    & 1.666     & 1.687                   \\
        \rowcolor{gray!20}
        \ALLMODELNAME  &     &\textbf{1.750}     &\textbf{1.541}   & \textbf{1.593}      & \textbf{1.562}      \\
        \midrule
        LS3DCG\cite{Habibie2021LearningS3} &\multirow{3}{*}{Show}     &2.692      & 2.807     &2.771    & 2.842      \\
        TalkSHOW\cite{talkshow:yi2022generating} &  &1.907     & 1.842   &1.742    &1.771     \\
        \rowcolor{gray!20}
        \ALLMODELNAME  &    & \textbf{1.330}                  & \textbf{1.278}                & \textbf{1.385}      & \textbf{1.321}     \\
        \midrule
    \end{tabular}
    }
       \vspace{-0.1in}
    \caption{\textbf{Results of the user study.} This shows the average rank of the user's favorite methods in terms of four metrics: Realism, Diversity, Gesture-Speech Synchrony and Semantic.}
    
    \label{tab:userstudymetric}
\end{table}

\subsection{Qualitative Results}
\mysubsub{Qualitative Comparisons}
We present qualitative comparisons (see in Fig.~\ref{fig:contrast.png}) on the BEATv2 dataset, comparing our method with GroundTruth (GT), EMAGE, and DiffuseStyleGesture. On the Show dataset (see in Fig.~\ref{fig:contrast2.png}), we compare our approach with LS3DCG and TalkSHOW. Our approach generates gestures that more closely resemble GT while maintaining better contextual semantic relevance.

\mysubsub{User Studies}
\label{sec4:user}
We conducted a user study on the BEATv2 and Show datasets, each with 10 video samples, recruiting 14 participants from diverse backgrounds. Participants rank different methods on four metrics: Realism (Real), Diversity (Div), Gesture-Speech Synchrony (G-S SYNC) and Semantic alignment (Sem). As shown in Tab.~\ref{tab:userstudymetric}, our method was consistently preferred across both datasets and all evaluation metrics, particularly excelling in realism and diversity. These results demonstrate that our MAG approach effectively generates lifelike, diverse, semantic-aware and rhythm-aware gestures that users enjoy.

\begin{table}[htbp]
    \vspace{0.08in}
\centering
\resizebox{0.95\linewidth}{!}{
\begin{tabular}{l|cccc}
\toprule
Method   & Rec$\downarrow$ & FGD$\downarrow$ & BC $\uparrow $&Diversity $\uparrow$ \\ \hline
VQ-VAE (Baseline) & 0.32 & 1.0830  & 6.1699  &  10.796        \\ 
VAE (wo VQ) & 0.06   & 0.0194     & 7.0836     &\textbf{13.032} \\
MT-VAE& 0.05   & 0.0194     & 7.0558     &\textbf{13.032} \\
MA-VAE& 0.05   & 0.0194     & 7.0891     &\textbf{12.922} \\
\VAENAME& 0.05  & \textbf{0.0174}     &\textbf{7.0837}     & 12.771 \\ 
\bottomrule
\end{tabular}
}
    \vspace{-0.1in}
\caption{\textbf{Ablation Studies on BEATv2 dataset}. We report metrics including Rec, FGD, BC $\times 10^{-1}$ and Diversity. The results show that the proposed MTA-VAE outperforms in generating more realistic, rhythm-aware gestures.}
\label{table:ablation_overall}
\end{table}



\begin{table}[htbp]
    \vspace{0.08in}
\centering
\resizebox{0.95\linewidth}{!}{
\begin{tabular}{l|ccc}
\toprule
Method    & FGD$\downarrow$ & BC $\uparrow$& Diversity$\uparrow$ \\ \hline
MAG (wo HGAT)  & 13.30  & 7.53  &  10.87        \\ 
MAG (w HGAT, wo Text) & 6.700     &7.71     & 12.25 \\ 
MAG (w HGAT) & \textbf{4.565}     &7.84     & 13.27 \\ 
\bottomrule
\end{tabular}}
    \vspace{-0.1in}
\caption{\textbf{Ablation Studies of HAGT.} We report metrics including FGD $\times 10^{-1}$, BC $\times 10^{-1}$ and Diversity. The results show effectiveness of HGAT.}
\label{table:ablation_overall2}
\end{table}


\subsection{Ablation studies}
\label{ab:stage2}


We report the ablation study on BEATv2 dataset and evaluate the impact of different variations of each proposed module by analyzing several key metrics: Reconstruction Loss of VAE (Rec),  Frechet Gesture Distance (FGD), Beat Consistency (BC), and Diversity (DIV). Rec metric evaluates the fidelity of the VAE model in terms of how accurately it can reconstruct the input data from its latent representations.

\mysubsub{Effect of Continuous Motion Representation} Tab.~\ref{table:ablation_overall} shows the comparison results with and without our continuous motion representation, where method ’VQ-VAE (Baseline)’ means baseline with discrete vector quantized motion tokenization, and ’VAE (wo VQ)’ means baseline without vector quantization, i.e., we represent motion features as continuous embeddings. 
The results show that ’VQ-VAE (Baseline)’ has the highest Rec and FGD, indicating poor reconstruction and a significant distribution gap between generated and real gestures due to information loss from quantization. In contrast, ’VAE (wo VQ)’ significantly improves Rec, FGD (from 1.0830 to 0.0194), BC, and Diversity by preserving the continuity of motion data, resulting in better reconstruction. 

\mysubsub{Effect of Contrastive Learning in Motion,Text and Audio}
Tab.~\ref{table:ablation_overall} shows the comparison results of contrastive learning with different modal features, where method ’MT-VAE’ means our continuous motion representation with contrastive learning in motion and text, ’MA-VAE’ means in motion and audio, and ’MTA-VAE’ means in motion, text and audio. 
The results show that 'MT-VAE'and 'MA-VAE' maintain similar improvements while further enhancing reconstruction quality. 'MTA-VAE' further maintains high BC and Diversity while preserving low FGD, highlighting the effectiveness of multi-modal learning in generating realistic and synchronized gestures. 
Furthermore, compared with method 'MAG (VAE)' of Tab.~\ref{tab:tab1}, method 'MAG (MTA-VAE)' of Tab.~\ref{tab:tab1} also proves the effectiveness of using contrastive learning in multi-modal features. 

\mysubsub{Effect of Hybrid Granularity Audio-Text Fusion Block (HGAT)}
Tab.~\ref{table:ablation_overall2} shows the comparison results of with and without HGAT, where method 'MAG (wo HGAT)' means our proposed MAG framework without HGAT, i.e., a typically MLP aligned fusion strategy is used to provide features to autoregressive transformer for fair comparison, 'MAG (w HGAT)' means MAG with our HAGT. The results show that HGAT can significantly improve the performance of all metrics, especially that FGD drops from 13.30 to 4.565, demonstrating the effectiveness of HGAT. Furthermore, 'MAG (w HGAT, wo Text)' in Tab.~\ref{table:ablation_overall2} means removing text feature extraction of HGAT, further prove the effectiveness of multi-modal learning.  

\section{Conclusion}
In this paper, we propose MAG, a novel approach for generating realistic co-speech gestures. Our method addresses challenges overlooked in previous works. Many existing approaches use autoregressive models with vector quantization to tokenize motion data, leading to information loss. We introduce MMAG, a framework that enables autoregressive modeling in continuous space, enhancing the quality and fluidity of generated gestures.  We also present MTA-VAE, which improves motion generation by aligning motion, audio, and text in latent space. Our approach achieves state-of-the-art performance on two public datasets, both quantitatively and qualitatively. The qualitative results and user study demonstrate that our method produces realistic, agile, and diverse gestures that are well-aligned with speech.


\clearpage
{\small
\bibliographystyle{ieee_fullname}
\bibliography{bibliography}
}


\end{document}